\documentclass[conference]{IEEEtran}
\IEEEoverridecommandlockouts
\usepackage{cite}
\usepackage{amsmath,amssymb,amsfonts}
\usepackage{algorithmic}
\usepackage{graphicx}
\graphicspath{{Figs/}}
\usepackage{textcomp}
\usepackage{xcolor}
\usepackage{url}
\usepackage{booktabs}
\usepackage{multirow}
\usepackage{siunitx}
\def\BibTeX{{\rm B\kern-.05em{\sc i\kern-.025em b}\kern-.08em
    T\kern-.1667em\lower.7ex\hbox{E}\kern-.125emX}}
\begin{document}

\title{Game AI Not Fun?\\ A Scoping Review and Meta-Analysis on the Differences in Enjoyment between Human and Computer Opponents
\thanks{This work was supported by Sony Interactive Entertainment Inc.}
}

\author{\IEEEauthorblockN{Ray Ito}
\IEEEauthorblockA{\textit{Graduate School of Interdisciplinary Information Studies,} \\
\textit{The University of Tokyo}\\
Tokyo, Japan \\
\url{https://orcid.org/0009-0001-7627-8178}
}
}

\IEEEoverridecommandlockouts
\IEEEpubid{\makebox[\columnwidth]{979-8-3315-9476-3\/26\/\$31.00 \copyright2026 IEEE \hfill}
\hspace{\columnsep}\makebox[\columnwidth]{ }}

\maketitle

\IEEEpubidadjcol

\begin{abstract}
Although advancements in game character AI aim to enhance player engagement, evidence suggests that perceiving an opponent as artificial can diminish the psychological experience. This paper presents a scoping review and meta-analysis of empirical studies focusing on player enjoyment when competing against human versus computer opponents. First, the scoping review was conducted to map the landscape of 20 included studies, detailing their study designs, outcome measures, and research foci. Second, a three-level meta-analysis synthesizing baseline comparisons from nine studies quantitatively assesses the differences in enjoyment. The results demonstrate a statistically significant, medium-to-large pooled effect size, indicating a psychological penalty in computer-opponent conditions. This paper provides a comprehensive overview of the extant knowledge on this topic, and underscores the necessity for further research in order to fully understand and resolve the penalty of the computer opponent context.
\end{abstract}

\begin{IEEEkeywords}
Games, AI, Context, Motivation, Psychology, Game Experience
\end{IEEEkeywords}

\section{Introduction}

The technology of game character AI is now advancing for better entertainment. Character AI is a category of game AI defined by Miyake~\cite{mcs} as an artificial intelligence that controls the behavior of in-game characters, such as opponents, teammates, or non-player characters (NPCs). Since the time of Pac-Man (Namco, 1980)~\cite{japanese_gameai_history}, developers and researchers have put efforts into improving character AI technology, employing the latest techniques, such as large language models~\cite{gpt_for_games}. These improvements aim to enhance player experience, as the ultimate goal of game AI is to elevate player enjoyment and engagement~\cite{game_ai_revisited}.

However, it remains doubtful whether, even if character AI acts exactly equivalent to humans, AI competitors can entertain as much as human competitors without reducing player experience. A review paper reported that the mere context of interacting with an AI diminished, no matter what the actual behavior was, users' prosociality, morality, and perceived likeability towards the opponent~\cite{AiContextReview}. Furthermore, a neuroscientific review paper reported that the brain responses to the presence of human and AI differ, including the rewarding system~\cite{AiContextNeuroReview}.

To bridge these general psychological findings and game-specific applications, organizing the fragmented empirical results regarding player enjoyment (which is often interpreted as intrinsic motivation in psychology~\cite{Ryan}) is important. The fact that the mere perception of whether interacting with a human or a computer alters the psychological experience is a critical concern not solely for psychologists but also for game developers and researchers aiming to enhance game character AI for better gaming experiences. Synthesizing this scattered knowledge through a scoping review effectively reveals practical design challenges for game AI practitioners and researchers, while simultaneously clarifying established findings and research gaps for the broader psychological community.

To address this issue, this paper presents a comprehensive scoping review and meta-analysis of empirical studies focusing on player enjoyment when competing against human and computer opponents. First, I systematically map the landscape of the related research, highlighting study designs, outcome measures, and research foci across 20 included studies. Second, a quantitative synthesis of baseline differences in enjoyment between human and computer opponent conditions through a three-level meta-analysis was conducted. These works together provide a comprehensive overview of the current state of knowledge on how opponent identity influences player enjoyment, offering insights for both theoretical understanding and practical applications in game design, game AI research, and psychological research.

\section{Study Selection}
\begin{figure}[htb]
    \centering
    \includegraphics[width=0.6\linewidth]{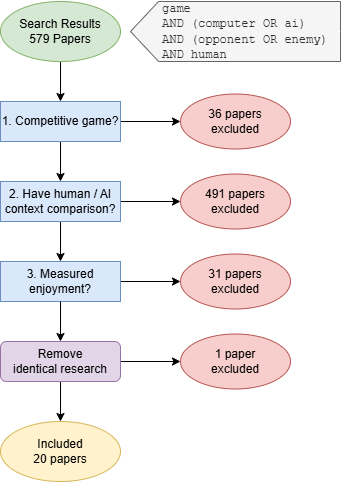}
    \caption{Flowchart of the paper screening process.}
    \label{fig:inclusion_flowchart}
\end{figure}

In order to collect papers of interest as comprehensively as possible, I employed TREE (UTokyo REsource Explorer) for paper collection, which is an instance of Ex Libris Summon provided by the University of Tokyo Library. TREE is a search engine that indexes a wide range of academic resources to which the University of Tokyo has access, including ProQuest, Web of Science, Elsevier, Springer, Taylor \& Francis, Wiley, Nature Publishing Group, IEEE, LexisNexis, PubMed, and more. This system is available online for everyone, not limited to the University of Tokyo members, which allows anyone to replicate the search process and results.

The search query was designed to capture papers that compare player psychological impact in the context of playing against human vs. computer opponents. The search string used was:
\begin{verbatim}
game AND (computer OR ai)
 AND (opponent OR enemy) AND human
\end{verbatim}

This search string was applied to the title, abstract, and keywords of indexed papers. In addition, ``Have full text\footnote{This refers to papers that the University of Tokyo has access to.}'', ``Academic'', ``Journal'', ``Proceedings'', and ``English'' filters were applied. The search was conducted on January 27, 2026, and it returned a total of 584 items. After removing duplicates, 579 unique papers were identified for screening.

The inclusion criteria for the scoping review were as follows:
\begin{enumerate}
    \item The paper is related to competitive games, where players compete against each other.
    \item The paper reports experiments comparing two conditions: one where participants believe they are playing against a human opponent, and another where they notice themselves playing against a computer opponent.
    \item The paper reports measurements of constructs related to enjoyment: self-reported enjoyment, emotional valence, motivation~\cite{Ryan}, flow~\cite{flow}, behavioral engagement, or neural rewarding.
\end{enumerate}

As shown in Fig.~\ref{fig:inclusion_flowchart}, 
based on the title and abstract screening, 36 papers were excluded by the first criterion, 491 papers were excluded by the second criterion, and 31 papers were excluded by the third criterion. There was a pair of papers reporting an identical experiment and its analysis~\cite{dup_345, reviewing_73}. Therefore, the one with more abundant content~\cite{reviewing_73} was included, and the other one~\cite{dup_345} was excluded. Finally, this process resulted in 20 papers included in this review~\cite{reviewing_12, reviewing_14, reviewing_20, reviewing_54, reviewing_73, reviewing_74, reviewing_95, reviewing_126, reviewing_127, reviewing_138, reviewing_224, reviewing_283, reviewing_366, reviewing_390, reviewing_412, reviewing_420, reviewing_457, reviewing_479, reviewing_515, reviewing_547}.

\section{Scoping Review}
This scoping review systematically mapped the existing literature on player enjoyment when competing against human versus computer opponents in order to offer readers a comprehensive overview of methodological standards, gaps, and opportunities for future research.

\subsection{Data Charting}
For each included study, data were systematically extracted across three primary domains. 
Regarding study designs, I recorded the game genre, the application of the Wizard of Oz designs~\cite{wizardofoz} (including the identity of the actual opponent controller), the participant allocation design (within- or between-subjects), and the specific instructional terminology used to describe the computer opponent. 
In terms of outcome measures, the measurement instruments and the targeted bio-signals were identified. This was limited to the measures directly related to enjoyment, motivation, and emotional valence and arousal, excluding other constructs, such as performance, targeted attitude, learning outcomes, or others.
Finally, the research foci were characterized through three distinct lenses: 
(1)~gradational manipulations of opponent characteristics, 
(2)~modulation by player attributes and environments, and 
(3)~modulations by task structures. 
For each perspective, the specific independent variables manipulated in the studies were documented.

\subsection{Study Designs}

Experimental frameworks of the 20 included studies were characterized across four primary dimensions: game genres, the application of the Wizard of Oz designs, participant allocation strategies, and the instructional framing of opponent identities.

\subsubsection{Game Genres}

Eleven studies utilized abstract or economic games, simplified, rule-governed experimental paradigms consisting of discrete, mathematically structured decision trials with minimal sensory-motor complexity. These included social dilemma and resource allocation tasks such as the Prisoner's Dilemma~\cite{reviewing_390, reviewing_457} and the Ultimatum or Dictator Games~\cite{reviewing_138, reviewing_366}. Other studies employed asymmetric or symmetric choice tasks like matching-pennies~\cite{reviewing_126, reviewing_515}, rock-paper-scissors~\cite{reviewing_547}, dominoes~\cite{reviewing_420, reviewing_479}, and basic logic or card games~\cite{reviewing_20, reviewing_74}. These paradigms isolate specific cognitive and neural mechanisms by minimizing extraneous visual and mechanical variables. 

Conversely, eight studies employed complex entertainment games which did not involve physical engagement. The genres utilized included first-person shooters~\cite{reviewing_14, reviewing_95, reviewing_127, reviewing_283}, racing game~\cite{reviewing_73}, action and puzzle games~\cite{reviewing_127, reviewing_283}, artillery tactical games~\cite{reviewing_54}, and role-playing games~\cite{reviewing_12}. Research in this category tends to evaluate under more ecologically valid condition for game design study.

Finally, two studies investigated exergames, focusing specifically on physical rehabilitation using an adaptive arm-movement game~\cite{reviewing_224} and exercise adherence using an online cycling exergame~\cite{reviewing_412}. In these contexts, the interaction with human versus computer opponents was assessed not merely for psychological enjoyment, but for its efficacy in sustaining physical effort and motivation over time.

\subsubsection{Employment of the Wizard of Oz Design}

Among the 20 included papers, 11 papers employed the Wizard of Oz design~\cite{reviewing_12, reviewing_14, reviewing_20, reviewing_54, reviewing_74, reviewing_138, reviewing_420, reviewing_457, reviewing_479, reviewing_515, reviewing_547}. Within this subset, only one paper~\cite{reviewing_14} had human-controlled opponents, and the rest had computer-controlled opponents. 

Conversely, the remaining nine papers did not employ the Wizard of Oz design, which means that the opponents were actually controlled by a human or a computer as participants believed. The reasons for the absence of the Wizard of Oz design were categorized into three: (1) the relationship between the participant and the opponent (i.e., if they know each other, or they are in the same room) was included as a dependent variable~\cite{reviewing_73, reviewing_95, reviewing_127, reviewing_283}, (2) the studies prioritized the ecological validity of the behavior of the opponents~\cite{reviewing_126, reviewing_366, reviewing_390}, and (3) the experiment duration was too long to maintain the deception~\cite{reviewing_224,reviewing_412}.

\subsubsection{Between- vs. Within-Subjects Designs}

Eleven studies adopted a within-subjects design, where the same participants were exposed to both human and computer opponent conditions~\cite{reviewing_14,reviewing_73,reviewing_95,reviewing_126,reviewing_127,reviewing_224,reviewing_283,reviewing_366,reviewing_420,reviewing_457,reviewing_515}. This approach is particularly prominent in neuroimaging and physiological studies. 

Conversely, seven studies utilized a between-subjects design, allocating participants to interact with either a human or a computer opponent, but not both~\cite{reviewing_12, reviewing_20,reviewing_54,reviewing_74,reviewing_138,reviewing_390,reviewing_412}.

Finally, the remaining two studies employed a mixed-design approach~\cite{reviewing_479,reviewing_547}. Both studies compared individuals with Autism Spectrum Disorder (ASD) and typically developing controls (between-subjects factor) while observing their interactions against different opponents (within-subjects factor), thereby isolating clinical differences in social motivation and mentalization.

\subsubsection{Instructional Framings of Opponent Identity}

The majority of the studies described the computer opponent as a blackbox. Eight studies employed the word ``computer'' to describe the computer opponent~\cite{reviewing_12,reviewing_14,reviewing_95,reviewing_127,reviewing_224,reviewing_283,reviewing_366,reviewing_515}, while three other studies used the term ``AI''~\cite{reviewing_20,reviewing_54,reviewing_412}.

Four studies visually presented the computer opponent. Of those four, two studies presented the opponents as virtual characters through monitors~\cite{reviewing_73,reviewing_390}, and the other two studies presented the opponents as physical robots~\cite{reviewing_138, reviewing_457}.

Five studies explicitly disclosed the algorithm's operating principles to the participants. Three studies described the computer opponent as random moves generators~\cite{reviewing_420,reviewing_479,reviewing_547}, two studies as a fixed algorithm~\cite{reviewing_74,reviewing_126}, and two studies as an adaptive learning algorithm~\cite{reviewing_126,reviewing_547}. The duplication of two studies~\cite{reviewing_126,reviewing_547} occurred because they had more than two experimental conditions.

The majority of the studies described the human opponent anonymously or without specific social attributes. Five studies simply employed the word ``human'' to describe the opponent~\cite{reviewing_14,reviewing_54,reviewing_138,reviewing_390,reviewing_457}, while four studies used the term ``another participant'' or fellow participants~\cite{reviewing_12,reviewing_20,reviewing_73,reviewing_74}. Four studies framed the opponent as the experimenters or the research assistants~\cite{reviewing_420,reviewing_479,reviewing_515,reviewing_547}. Four studies labeled the opponent as a ``stranger'' or ``real players''~\cite{reviewing_95,reviewing_127,reviewing_283,reviewing_412}, and one study used specific personal names to represent real human characters~\cite{reviewing_366}. 

Seven studies provided specific social attributes or roles to the human opponent. Three studies described them as a member of a specific peer group, such as a student from the same university, an outgroup student, or another clinical patient~\cite{reviewing_54,reviewing_126,reviewing_224}. Four studies explicitly situated the opponent within the participant's existing interpersonal networks. All of these four studies used the term ``friend'' or ``relative'' to frame the human opponent~\cite{reviewing_95,reviewing_127,reviewing_224,reviewing_283}. The duplication of several studies~\cite{reviewing_54,reviewing_95,reviewing_127,reviewing_224,reviewing_283} across these categories is because they included multiple human-opponent experimental conditions within their designs.

\subsection{Outcome Measures}

To capture the multifaceted nature of enjoyment when competing with human versus computer opponents, the included studies employed a diverse array of measurement methodologies. These approaches were categorized into four modalities: self-reported, behavioral, neuroimaging, and physiological measures.

\subsubsection{Self-Reported Measures}
Twelve studies utilized subjective questionnaires to assess enjoyment, motivation, engagement, and emotional states. 

Seven papers employed standardized instruments. Standardized questionnaires regarding motivation included the Flow Short Scale~\cite{reviewing_12, reviewing_73}, the Intrinsic Motivation Inventory~\cite{reviewing_224}, the Game Experience Questionnaire~\cite{reviewing_14}, and the ITC-Sense of Presence Inventory~\cite{reviewing_127, reviewing_283}. To measure emotional valence and arousal, studies adopted the Self-Assessment Manikin~\cite{reviewing_127, reviewing_283} and the Affect Grid~\cite{reviewing_138}. 

On the other hand, seven studies used original scales, which can be categorized into two primary domains based on their target constructs: direct enjoyment and fun~\cite{reviewing_12, reviewing_54, reviewing_74, reviewing_224, reviewing_457}, and engagement and motivation (e.g., the explicit desire to win)~\cite{reviewing_420, reviewing_479}.

\subsubsection{Behavioral Measures}
Four studies captured objective behavioral indicators as representative of motivation. Two studies reported binary choice to voluntarily re-play or re-challenge the opponent after a match~\cite{reviewing_20, reviewing_74}. Two studies measured longitudinal motivation through the continuous logging of play duration and frequency across multiple sessions~\cite{reviewing_224, reviewing_412}. Additionally, one study working within an exergame context measured physical exertion by analyzing inactive play duration~\cite{reviewing_412}.

\subsubsection{Neuroimaging Measures}
Seven studies utilized functional magnetic resonance imaging (fMRI) to investigate the neural correlates of reward processing and social cognition during gameplay. The ventral striatum and nucleus accumbens were the most prominently examined regions, measured in five studies to capture reward prediction errors and differential responses to winning or gaining points~\cite{reviewing_14, reviewing_126, reviewing_366, reviewing_420, reviewing_479}. Other targeted regions included the ventromedial prefrontal cortex (vmPFC) and orbitofrontal cortex for reward network activity~\cite{reviewing_14, reviewing_126, reviewing_420}, the dorsal striatum (caudate and putamen)~\cite{reviewing_14, reviewing_515}, the midbrain~\cite{reviewing_515}, and the hypothalamus for social motivation~\cite{reviewing_547}.

\subsubsection{Physiological Measures}
Five studies collected bio-signals and automated facial expression data to objectively assess continuous emotional valence and autonomic arousal. Facial electromyography (EMG) targeting the zygomaticus major, orbicularis oculi, and corrugator supercilii was employed in three studies to index positive and negative affective reactions to in-game events~\cite{reviewing_95, reviewing_127, reviewing_283}. Autonomic arousal was quantified using electrodermal activity (EDA/SCL)~\cite{reviewing_95, reviewing_127} and electrocardiography (ECG/HR)~\cite{reviewing_127, reviewing_283, reviewing_412}. Furthermore, one study applied the FACET automated expression analysis tool to extract synchronized emotional intensities from webcam recordings~\cite{reviewing_390}.

\subsection{Study Foci}

The research foci of the included studies are categorized into three primary analytical lenses to identify the specific variables influencing player experience. These include: (1) gradational manipulations of opponent characteristics, (2) the moderating effects of player attributes and environmental contexts, and (3) the influence of specific task structures.

\subsubsection{Gradational Manipulations of Opponent Characteristics}
Beyond the binary comparison of human versus computer opponents, seven studies manipulated the opponent's characteristics in a gradational manner. These studies can be categorized by the specific dimensions manipulated to explore nuanced research questions.

\paragraph{Social Relationship}
Four studies by the same research group manipulated the social relationship and co-location of the opponent, comparing a computer baseline with a human stranger and a human friend~\cite{reviewing_95, reviewing_127,reviewing_224, reviewing_283}. They investigated how the opponent's social identity and physical proximity influence spatial presence, physiological arousal, and momentary emotional reactions during gameplay.

\paragraph{Algorithmic Sophistication}
One study isolated the artificial opponent's perceived intelligence and adaptability~\cite{reviewing_126}. By comparing a human opponent with a ``fixed'' algorithm following a Nash equilibrium and a ``learning'' algorithm adapting to the player, this study aimed to uncover the distinct neural correlates of second-order strategic inferences and human agency perception.

\paragraph{Anthropomorphic Appearance}
Two studies systematically manipulated the physical appearance and embodiment of non-human opponents~\cite{reviewing_138, reviewing_457}. By comparing human partners with artificial opponents ranging from a standard computer to functional and highly anthropomorphic robots, they explored how the degree of human-likeness alters anthropomorphic attributions, economic behaviors, and the recruitment of mentalizing brain regions.

\paragraph{Hybrid Manipulation of Appearance and Intelligence}
One study employed a hybrid manipulation combining embodiment and algorithmic sophistication, comparing interactions against a human, an AI-equipped robot, and a computer acting randomly~\cite{reviewing_547}.

\subsubsection{Modulation by Player Attributes and Environments}
Several studies examined how individual player attributes and environments modulate the experience of playing against different opponents.

\paragraph{Clinical Differences} Two studies compared individuals with Autism Spectrum Disorder (ASD) to typically developing controls, focusing on neural activations related to mentalizing and social motivation when facing human versus artificial agents~\cite{reviewing_479, reviewing_547}.

\paragraph{Demographic Differences}
One study conducted separate experiments whose participants differed by demographic factors. Yokoi and Nakayachi~\cite{reviewing_74} conducted three experiments: two in Japan and one in the United States.

\paragraph{Psychological Characteristics}
Two studies employed psychological characteristic assessments as independent variables. Ravaja~\cite{reviewing_127} measured behavioral activation/inhibition system sensitivity (sensitivity towards rewards and punishments), impulsivity, and sensation seeking traits of the participants to investigate their moderating effects. Williams~\cite{reviewing_73} treated prior videogame interest and weekly play hours as continuous variables to determine their correlation with intrinsic engagement and performance.

\subsubsection{Modulations by Task Structures}
Five studies used winning or losing events as independent variables~\cite{reviewing_14, reviewing_95, reviewing_126, reviewing_420, reviewing_479}. Brüne et al.~\cite{reviewing_366} manipulated the opponent's prior fairness in economic games, Takahashi et al.~\cite{reviewing_515} controlled the pre-determined winning probability across blocks, and Stiff and Kedra~\cite{reviewing_54} manipulated the existence of a cooperative partner.

\section{Baseline Meta-Analysis}
In order to quantitatively synthesize the baseline enjoyment differences between human and computer opponent conditions across the included studies, a meta-analysis was conducted. Baseline difference was defined as the pure comparison between the human-opponent and computer-opponent conditions without any additional manipulations on participants or opponents. This meta-analysis provides an estimate of the overall effect size of opponent identity on player motivation, arousal, and valence while accounting for the heterogeneity across studies in terms of constructs measured and modalities used.

\subsection{Methods}
\subsubsection{Comparison Selection}
The comparison considered as the baseline difference was selected based on the following criteria: (1)~it did not involve any additional manipulations of opponent characteristics, player attributes, or task structures, (2)~paper reported numerical data (excluding fMRI data) of the outcome measures related to motivation, arousal, and valence, and (3)~no data duplication with another paper.

\subsubsection{Effect Size Computation}
All comparisons were converted to Hedges' $g$~\cite{hedges1981} via
three pathways depending on available statistics: (1)~pooled-SD estimator
(SMD) or standardized mean change with correlation correction (SMCC;
$r = 0.50$ assumed) for continuous outcomes, rescaled to a common
between-subjects scale as $d_{\text{between}} = d_z\sqrt{2(1-r)}$
\cite{morris2008,nakagawa2007}; (2)~log odds ratio converted to SMD via
$g = \text{LOR}\cdot\sqrt{3}/\pi$~\cite{hasselblad1995}; and
(3)~$d_z = \sqrt{F/N}$ for within-subjects $F$-statistics with
$df_\text{effect}=1$~\cite{nakagawa2007}.
All analyses used \texttt{metafor}~\cite{viechtbauer2010}.

\subsubsection{Three-Level Model}
Following  Van den Noortgate et al.~\cite{vandennoorgate2013}, a three-level REML model was specified to
account for multiple effect sizes per study:
\begin{equation}
  y_{ij} = \mu + u_j + v_{ij} + \varepsilon_{ij},
\end{equation}
where $u_j \sim \mathcal{N}(0,\sigma^2_3)$,
$v_{ij} \sim \mathcal{N}(0,\sigma^2_2)$, and
$\varepsilon_{ij} \sim \mathcal{N}(0, v_{ij})$~\cite{cheung2014}.
Fit was compared to a two-level model via LRT (ML estimation).
Moderators were tested via omnibus $Q_M$ tests.
Sensitivity to assumed $r$ was assessed at $r \in \{0.3, 0.5, 0.7\}$.

\subsection{Results}

\subsubsection{Selected Comparisons}
Across 20 included studies, 25 comparisons in nine studies~\cite{reviewing_12, reviewing_14, reviewing_20, reviewing_73, reviewing_74, reviewing_127, reviewing_420, reviewing_479, reviewing_283} met the criteria for baseline differences. The data in the remaining 11 studies were not included in the meta-analysis for the following reasons: (1)~all participants were characterized in two papers~\cite{reviewing_224, reviewing_412}, (2.1)~numerical data required for effect size calculation was not reported in four papers~\cite{reviewing_54,reviewing_138, reviewing_390,reviewing_457}, (2.2)~outcome of interest was only measured by fMRI in four papers~\cite{reviewing_126,reviewing_366,reviewing_515,reviewing_547}, and (3)~the duplicated data were used in one paper~\cite{reviewing_95} (because Kivikangas and Ravaja~\cite{reviewing_95} analyzed the identical data with Ravaja et al.~\cite{reviewing_283}, only Ravaja et al.~\cite{reviewing_283} was included in the meta-analysis).

\subsubsection{Included Effect Sizes}

\begin{table*}[t]
\caption{Individual Hedges' $g$ Effect Sizes Included in the Meta-Analysis.}
\label{tab:effect_sizes}
\centering
\small

\begin{tabular}{@{}lllcS[table-format=-1.3]S[table-format=1.3]lc@{}}
\toprule
Study & Construct & Modality & Design~$^1$ &
{Hedges' $g$} & {$SE$} & {95\% CI} & Source~$^2$ \\
\midrule
\cite{reviewing_73}  & Motivation & Questionnaire & W &  0.242 & 0.103 & $[0.04,\;0.44]$ & Cont. \\
\midrule
\cite{reviewing_12}  & Motivation & Questionnaire & B &  0.592 & 0.244 & $[0.11,\;1.07]$ & Cont. \\
                     & Motivation & Questionnaire & B &  0.612 & 0.245 & $[0.13,\;1.09]$ & Cont. \\
\midrule
\cite{reviewing_14}  & Valence    & Questionnaire & W &  0.831 & 0.281 & $[0.28,\;1.38]$ & Cont. \\
                     & Valence    & Questionnaire & W &  1.100 & 0.307 & $[0.50,\;1.70]$ & Cont. \\
\midrule
\cite{reviewing_20}  & Motivation & Behavioral    & B &  0.594 & 0.156 & $[0.29,\;0.90]$ & Binary \\
\midrule
\cite{reviewing_74}  & Motivation & Behavioral    & B & -0.053 & 0.306 & $[-0.65,\;0.55]$ & Binary \\
                     & Motivation & Behavioral    & B &  0.188 & 0.135 & $[-0.08,\;0.45]$ & Binary \\
                     & Motivation & Behavioral    & B &  0.052 & 0.201 & $[-0.34,\;0.45]$ & Binary \\
\midrule
\cite{reviewing_127} & Valence    & Questionnaire & W &  1.825 & 0.340 & $[1.16,\;2.49]$ & Cont. \\
                     & Arousal    & Questionnaire & W &  0.644 & 0.229 & $[0.19,\;1.09]$ & Cont. \\
                     & Valence    & Facial EMG    & W &  0.718 & 0.234 & $[0.26,\;1.18]$ & Cont. \\
                     & Valence    & Facial EMG    & W &  1.517 & 0.313 & $[0.90,\;2.13]$ & Cont. \\
                     & Arousal    & Facial EMG    & W &  2.498 & 0.423 & $[1.67,\;3.33]$ & Cont. \\
                     & Arousal    & EDA/SCL       & W &  0.623 & 0.233 & $[0.17,\;1.08]$ & Cont. \\
                     & Arousal    & ECG/IBI       & W &  0.772 & 0.238 & $[0.31,\;1.24]$ & Cont. \\
                     & Motivation & Questionnaire & W &  0.965 & 0.252 & $[0.47,\;1.46]$ & Cont. \\
\midrule
\cite{reviewing_283} & Valence    & Questionnaire & W &  0.836 & 0.202 & $[0.44,\;1.23]$ & $F$-stat \\
                     & Valence    & Facial EMG    & W &  0.564 & 0.200 & $[0.17,\;0.96]$ & $F$-stat \\
                     & Valence    & Facial EMG    & W &  1.685 & 0.305 & $[1.09,\;2.28]$ & $F$-stat \\
                     & Arousal    & Facial EMG    & W &  2.323 & 0.393 & $[1.55,\;3.09]$ & $F$-stat \\
                     & Arousal    & ECG/IBI       & W &  0.899 & 0.210 & $[0.49,\;1.31]$ & $F$-stat \\
                     & Motivation & Questionnaire & W &  0.718 & 0.195 & $[0.34,\;1.10]$ & $F$-stat \\
\midrule
\cite{reviewing_420} & Motivation & Questionnaire & W &  0.247 & 0.239 & $[-0.22,\;0.72]$ & Cont. \\
\midrule
\cite{reviewing_479} & Motivation & Questionnaire & W &  0.282 & 0.273 & $[-0.25,\;0.82]$ & Cont. \\
\bottomrule
\end{tabular}

\begin{minipage}{14cm}
$^1$ W\,=\,within-subjects; B\,=\,between-subjects.

$^2$ How the data were obtained: Cont.\,=\,continuous (mean\,+\,SD);
Binary\,=\,2$\times$2 frequency table;
$F$-stat\,=\,derived from $F$-value and $\eta^2_p$.
Positive $g$ indicates higher response in the human-opponent condition.
\end{minipage}

\end{table*}

A total of $k = 25$ effect sizes were extracted from nine independent studies.
Fifteen effects derived from continuous outcomes with reported means and SDs,
six from $F$-statistics with $\eta^2_p$, and four from binary re-challenge
frequencies. Table~\ref{tab:effect_sizes} lists each effect size with its
Hedges' $g$, standard error, 95\% confidence interval (CI),
measurement modality, and source pathway.

\subsubsection{Overall Effect}

The three-level REML model yielded a statistically significant pooled effect
of $g = 0.631$ (95\% CI $[0.319,\;0.943]$, $z = 3.96$, $p <.0001$),
indicating a medium-to-large advantage of human-opponent conditions over
computer-opponent conditions across all constructs and modalities.

Substantial heterogeneity was present ($Q(24) = 129.06$, $p <.0001$).
The variance was decomposed into two components: between-study variance
$\hat{\sigma}^2_3 = 0.124$ ($\sqrt{\hat{\sigma}^2_3} = 0.352$) and
within-study, between-effect variance
$\hat{\sigma}^2_2 = 0.148$ ($\sqrt{\hat{\sigma}^2_2} = 0.385$).
The LRT confirmed that the three-level structure provided a significantly
better fit than a two-level model
($\Delta\chi^2(1) = 10.07$, $p = .0015$), indicating that within-study
clustering of effect sizes was non-trivial and that ignoring it would
inflate precision~\cite{vandennoorgate2013}.

\subsubsection{Construct-Wise Sub-Analyses}

\begin{table}[htb]
\caption{Construct-Wise Pooled Effect Sizes from Multi-Level Meta-Analyses.
}
\label{tab:construct}
\centering
\small
\begin{tabular}{@{}lccS[table-format=1.3]lS[table-format=1.3]@{}}
\toprule
Construct & $k$ & $n_s$ & {$g$} & 95\% CI & {$p$} \\
\midrule
Arousal$^\dagger$   & 6  & 2 & 1.234 & $[0.58,\;1.89]$ & {$<$~.001} \\
Valence$^\dagger$   & 8  & 3 & 1.087 & $[0.76,\;1.41]$ & {$<$~.001} \\
Motivation          & 11 & 8 & 0.436 & $[0.22,\;0.65]$ & {$<$~.001} \\
\bottomrule
\end{tabular}

\begin{minipage}{7.5cm}
$k$\,=\,number of effect sizes; $n_s$\,=\,number of studies. \\
$^\dagger$Boundary solution: $\hat{\sigma}^2_3 \approx 0$ in three-level model;
two-level model fitted instead ($\hat{\sigma}^2_2$ not estimable separately).
\end{minipage}

\end{table}

Table~\ref{tab:construct} reports construct-specific pooled estimates.
All three constructs showed statistically significant effects.
Arousal and valence exhibited
large-to-very-large effects ($g \geq 1.09$), while the motivation construct,
which was represented by the greatest number of studies ($k = 11$, 8 studies),
yielded a moderate estimate ($g = 0.44$, 95\% CI $[0.22,\;0.65]$).
Arousal and valence estimates derive from two-level fallback models
(boundary solution; see Table~\ref{tab:construct}) and should be
interpreted with caution given the small number of studies.

\subsubsection{Moderator Analyses}

\begin{table}[t]
\caption{Omnibus Moderator Tests.}
\label{tab:moderators}
\centering
\small
\begin{tabular}{@{}lS[table-format=2.2]cS[table-format=1.3]@{}}
\toprule
Moderator & {$Q_M$} & $df$ & {$p$} \\
\midrule
Construct        & 14.86 & 2 & .001 \\
Modality         & 16.88 & 4 & .002 \\
Design type      &  3.31 & 1 & .069 \\
Effect source    &  3.90 & 2 & .142 \\
\bottomrule
\end{tabular}
\end{table}

Table~\ref{tab:moderators} presents omnibus $Q_M$ tests for each moderator.
Measured construct ($Q_M(2) = 14.86$, $p <.001$) and measurement modality
($Q_M(4) = 16.88$, $p = .002$) significantly moderated effect magnitude,
confirming that the human-opponent advantage is larger for physiological arousal
and valence indices than for self-report motivation. Experimental design type
(within vs.\ between subjects) showed a borderline non-significant trend
($Q_M(1) = 3.31$, $p = .069$), which may partly reflect the rescaling of
within-subjects effects using the assumed correlation $r = 0.50$.
Critically, the effect-size source (continuous, binary, or $F$-statistic) was
non-significant ($Q_M(2) = 3.90$, $p = .142$), indicating that the three
conversion pathways did not introduce systematic bias into the pooled
estimate~\cite{viechtbauer2010}.

\subsubsection{Sensitivity Analysis}

\begin{table}[t]
\caption{Sensitivity Analysis: Effect of Assumed Within-Participant Correlation $r$
on the Pooled Estimate.}
\label{tab:sensitivity}
\centering
\small
\begin{tabular}{@{}cS[table-format=1.3]S[table-format=1.3]lc@{}}
\toprule
$r$ & {$g$} & {$SE$} & {95\% CI} & {$p$} \\
\midrule
0.3 (conservative) & 0.717 & 0.189 & $[0.35,\;1.09]$ & {$<$~.001} \\
0.5 (baseline)     & 0.631 & 0.159 & $[0.32,\;0.94]$ & {$<$~.001} \\
0.7 (liberal)      & 0.524 & 0.124 & $[0.28,\;0.77]$ & {$<$~.001} \\
\bottomrule
\end{tabular}
\end{table}

Table~\ref{tab:sensitivity} reports the main model re-estimated under three
assumptions for the within-participant correlation $r$.
Across the full plausible range ($r = 0.3$ to $r = 0.7$), the pooled estimate
varied between $g = 0.52$ and $g = 0.72$, and remained highly significant
($p <.001$) in all scenarios. This indicates that the conclusion of a
significant medium-to-large human-opponent advantage is robust to uncertainty
about the within-subjects correlation assumed during effect-size scaling.

\section{Discussion}

This paper employed two complementary methodological approaches to synthesize 20 pieces of empirical literature on how opponent identity influences player enjoyment in competitive games. First, a scoping review systematically mapped the landscape of 20 empirical studies, characterizing their experimental designs, outcome measures, and research foci to provide a structured overview of what has been studied, how it has been measured, and where the gaps remain. Second, a three-level random-effects meta-analysis was conducted on 25 baseline effect sizes drawn from nine of those studies, yielding a pooled estimate which indicates a medium-to-large and statistically robust advantage of human-opponent conditions over computer-opponent conditions across motivational constructs and measurement modalities.

\subsection{Bias in the Included Studies}

The employed games had three genre categories: 11 abstract or economic games, 8 entertainment video games, and 2 exergames. This distribution may raise concerns about the generalizability towards practical game designs. While controllable abstract games provide precise isolation of cognitive and neural mechanisms, entertainment games provide a more ecologically valid context for evaluating the gaming experience in real-world game designs. Because entertainment games include a variety of forms and psychological noise, more studies involving more game instances are needed to draw generalizable conclusions, not limited to specific game designs, for practical applications.

Given that this paper focuses on the opponent context itself, not the behavior of the opponent, the use of the Wizard of Oz design is preferred to identify the contextual effects. In this review, 11 out of 20 studies employed the Wizard of Oz design, but most of the studies relied on computer behavior, and only one study had human-controlled opponents. Most studies conducted a control verification to confirm that the participants were deceived, but the absence of human-controlled opponents may limit the ecological validity of the findings, given the recent technological advances in game AI behaviors.

Thirteen studies adopted within-subjects designs regarding the opponent context, and seven between-subjects designs, which did not show a large bias.

With respect to the instructional framing of opponent identity, the majority of the studies described the computer opponent as ``computers,'' whereas only a limited number have used the term ``AI.'' In recent years, the term ``AI'' has become more prevalent in popular discourse and may carry different connotations. Therefore, investigating the contextual effects of the term ``AI'' now could be valuable.

The measurement instruments, especially the self-reported questionnaires, were highly heterogeneous across the included studies. Although some studies employed standardized instruments, many used original scales with varying constructs and items. This heterogeneity limits the comparability of results across studies.

Considerable research attempted to discover the root cause of the motivational differences between human and computer opponents by manipulating the opponent's characteristics: social relationship, algorithmic adaptability, and anthropomorphic appearance. However, these may not be the only factors that contribute to the motivational differences, and more diverse manipulations are needed to discover the underlying mechanisms of the contextual penalty of computer opponents. There also could be important factors attributable to players themselves. Yokoi and Nakayachi~\cite{reviewing_74}, the only study that directly compared demographic differences, found there was a significant difference between Japanese and American participants, with the former showing a lower motivation for computer opponents. Further investigation focusing on cultural differences (including differences by era) may be necessary as well.

\subsection{Game AI Not Fun?}

The baseline meta-analysis indicated that the mere context of playing against a computer opponent causes a significant penalty in motivation, arousal, and valence compared to playing against a human opponent.

Although this finding has shown the importance of the opponent context, it does not necessarily mean that game character AI is `useless.' As mentioned in the previous section, the included studies have not fully explored the various factors that may contribute to the motivational differences. For example, the context of strength, presentation of the opponent, and the narrative framing of the game may all modulate the motivational differences. Therefore, future research investigating the root causes of the motivational differences does not only have theoretical significance for pure psychology but also practical importance for game design and game AI research, as it can provide actionable insights for how to design more engaging and motivating game AI opponents.

\subsection{Limitations}

Both the scoping review and the meta-analysis have limitations. The scoping review may have missed relevant studies due to the search terms used. For instance, the psychological evidence of differences in enjoyment between human and computer competitors may not be limited to games, but also include other competitive tasks. Also, since the interactions between players and character AIs are not limited to competitive forms but also include cooperative scenarios, further investigation on cooperative contexts is required to observe the full spectrum of contextual effects of character AIs. Moreover, the search strategy itself may have missed relevant studies because it is limited to papers available to a specific university, and to English papers.

The meta-analysis is limited by the small number of studies and effect sizes included, as each individual study had a large influence on the pooled estimate. There may also be publication bias in this meta-analysis since the included studies were published in English journals or conferences. In addition, fMRI studies were excluded from the meta-analysis. These fMRI results should be synthesized using image-based methods such as CBMA or IBMA~\cite{Peraza2025} in the future for comprehensive neuroscientific implications.

\section{Conclusion}
This paper presented a scoping review and a meta-analysis examining the differences in player enjoyment when competing against human versus computer opponents. The scoping review offered a comprehensive overview of the existing literature, and the findings by meta-analysis revealed a statistically robust, medium-to-large penalty for playing against computer opponents compared to human opponents. This highlights the importance of opponent identity as a contextual factor influencing player experience. However, the underlying mechanisms driving these differences remain unclear, and further research is required to explore the various factors not only for theoretical understanding but also for practical implications in game AI design and development.

I hope this paper helps the interdisciplinary community encompassing psychologists, engineers, and practitioners, resulting in more fun game AI experiences for all players.

\section*{Acknowledgements}

This work was advised by Mr.~Tomohiro Oto\footnote{Sony Interactive Entertainment Inc.\label{sie}}, Mr.~Kentaro Suzuki\footref{sie}, Mr.~Yabe Hiroyuki\footref{sie}, Dr.~Youichiro Miyake\footnote{Institute of Industrial Science, The University of Tokyo}, and Dr.~Yusuke Mori.

\bibliographystyle{IEEEtran}
\bibliography{ref}

\end{document}